\documentclass[twocolumn]{revtex4-2}

\usepackage{amsmath}
\usepackage{float}
\usepackage{graphicx}
\usepackage{mathtools}
\DeclarePairedDelimiter{\ceil}{\lceil}{\rceil}
\usepackage{pgfplots}
\usepackage{ragged2e}
\usepackage{booktabs}
\usepackage{nicematrix,tikz}
\usetikzlibrary{decorations.pathreplacing}
\newlength\fwidth

\DeclareMathOperator{\sinc}{sinc}

\usepackage{xcolor}

\newcommand \bd{\color{blue}}

\usepackage{soul}

\begin{document}
		\title{The 3D coarse-graining formulation of interacting elastohydrodynamic filaments and multi-body  microhydrodynamics}
	\author{Paul Fuchter}
	\author{Hermes Bloomfield-Gad\^elha}
\email[]{hermes.gadelha@bristol.ac.uk}
\affiliation{Department of Engineering Mathematics and Bristol Robotics Laboratory, University of Bristol, Bristol, UK.}

\begin{abstract}
Elastic filaments are vital to biological, physical and engineering systems, from cilia driving fluid in the lungs to artificial swimmers and micro-robotics. Simulating slender structures requires intricate balance of elastic, body, active, and hydrodynamic moments, all in three-dimensions. Here, we present a generalised 3D coarse-graining formulation that is efficient, simple-to-implement, readily extendable and usable for a wide array of applications. Our method allows for simulation of collections of 3D elastic filaments, capable of full flexural and torsional deformations, coupled non-locally via hydrodynamic interactions, and including multi-body microhydrodynamics  of structures with arbitrary geometry. The method exploits the exponential mapping of quaternions for tracking three-dimensional rotations of each interacting element in the system, allowing for computation times up to 150 times faster than a direct quaternion implementation. Spheres are used as a `building block' of both filaments and solid micro-structures for straightforward and intuitive construction of arbitrary three-dimensional geometries present in the environment. We highlight the strengths of the method in a series of non-trivial applications including bi-flagellated swimming, sperm-egg scattering, and particle transport by cilia arrays. Applications to lab-on-a-chip devices, multi-filaments, mono-to-multi flagellated microorganisms, Brownian polymers, and micro-robotics are straightforward. A Matlab code is provided for further customization and generalizations.

\end{abstract}
	\maketitle
	\section{Introduction}
	
\begin{figure*}[t!]
    \includegraphics[width=\textwidth]{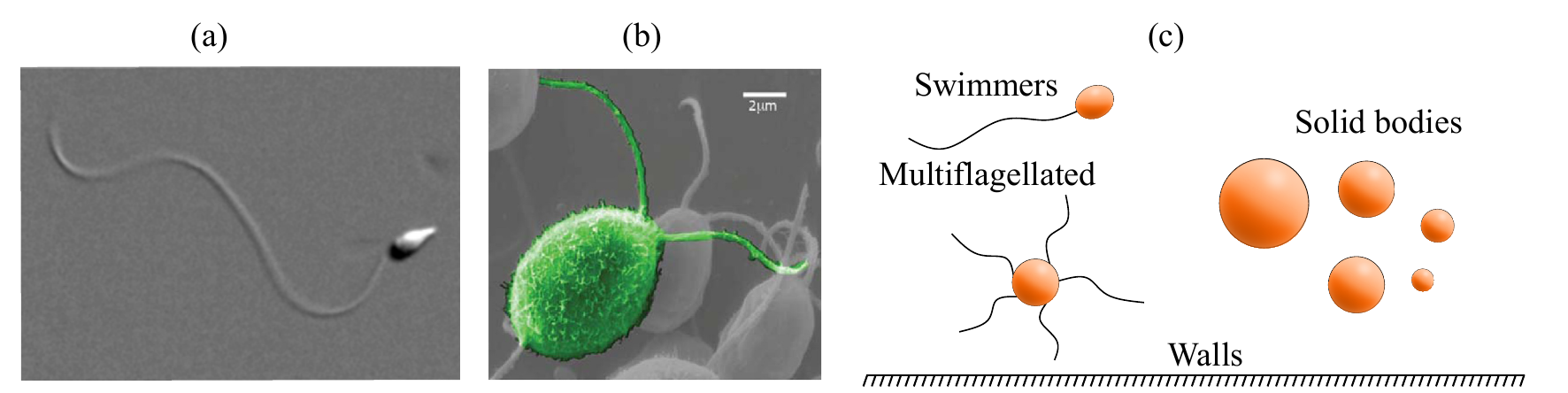}
    \caption{(a) Uni-flagellated sea urchin sperm~\cite{gong_steering_2020}. (b) Multiflagellated green algae \textit{Chlamydomonas}~\cite{elgeti_physics_2015}. (c) Representation of interactions that the method is capable of simulating: multiple filaments, filaments connected to arbitrary solid bodies, multi-filament-body structures and micro-structures with arbitrary shapes, free or fixed.}
    \label{intro_fig}
\end{figure*}

	Elastic passive and active filaments are the building blocks of numerous biological, physical, engineering and robotic systems. These include, but are not limited to, polymer and fibre dynamics, cilia driving fluid in the lungs, microtubules regulating deformations in cancerous cells, flagella propelling spermatozoa and algae (Fig.~\ref{intro_fig}(a,b)), and robotic micro-swimmers in microfluidic devices, among many others ~\cite{du_roure_dynamics_2019,gaffney_mammalian_2011,lauga_hydrodynamics_2009,cartwright_fluid-dynamical_2004,elgeti_physics_2015}. Simulating these structures fully in three-dimensions is challenging due to the convoluted balance of several interacting components. Moments arise from elasticity, internal activity, contact and the inertialess fluid that the filaments are embedded in. This complexity is augmented by the `non-local' presence of other structures embedded in the fluid, such as walls and solid bodies. Numerous computational architectures have been developed to date attempting to simulate these systems \cite{ryan_finite_2022,koshakji_robust_2023,schoeller_methods_2021,PhysRevFluids.6.014102,PhysRevFluids.7.074101,moreau_asymptotic_2018}, a testimony of the growing importance of filament fluid-structure interactions to the scientific community at large, and we direct the reader to the recent review on the topic by~\cite{du_roure_dynamics_2019} and references therein.

A hierarchy of fluid and elastic theories exist from continuous to discrete, with varying levels of complexity depending upon the precision required. Methods for resolving inertialess hydrodynamics include local theories, such as resistive force theory (RFT) \cite{gray_propulsion_nodate}, and more complex non-local theories, such as slender-body theories (SBT), regularised Stokeslets \cite{cortez_regularized_2018} and the Rotne-Prager-Yamakawa (RPY) approximation \cite{wajnryb_generalization_2013}, and boundary-element methods (BEM) \cite{pozrikidis_shear_2010}. Similarly, elastic deformation of filaments may be described using Cosserat rod theory, which accommodates bending, twist, shear and extensibility in three-dimensions \cite{cao_nonlinear_2008},  Kirchoff rod approximation \cite{nizette_towards_1999}, where filaments can only bend and twist, or even constraining inextensible filaments to planar deformations (with zero twist)~\cite{wiggins_flexive_1998}. Immersed boundary methods exploit direct numerical solution of Navier-Stokes equations with a Lagrangian mesh for the filament \cite{fauci_computational_1988}, whilst in bead models the filament is discretised into segments, and the elastic coupling is described via discrete `elastic bonds' \cite{elgeti_hydrodynamics_2010,delmotte_general_2015}, among many others.

	Classical formulations of solving filament elastohydrodynamics revolves around the continuum limit of moment balance~\cite{tornberg_simulating_2004,gadelha_nonlinear_2010}. They are critical for analytical progress and direct model interpretability from the resulting partial differential equations (PDEs)~\cite{wiggins_flexive_1998}. However, the PDEs have a high-order, are highly nonlinear, and are coupled with a boundary value problems (BVP), which require numerous boundary conditions. Unsurprisingly, these equations are challenging to solve even in lower dimensions~\cite{gadelha_nonlinear_2010,gadelha_flagellar_2019}. They are also numerically stiff and thus computationally expensive/time-consuming, often requiring penalization strategies to regularise numerical errors~\cite{tornberg_simulating_2004}. They also require specific numerical architecture that is not easily transferable among different applications, especially in 3D, nor easily generalizable to systems involving non-trivial geometry and/or possessing many, and distinct, interacting units, such as elastic filaments interacting with solid structures (Fig.~\ref{intro_fig}(c)). As we shall see below, the formulation derived here will attempt to resolve these challenges.  
		
	Recent developments in coarse-graining formulations offer intuitive model interpretability, straightforward implementation, and numerical  advantages over previous methods. By asymptotically integrating the momentum density over `coarse' segments along the filament, the coarse-graining method arises from first principles, and removes the need to solve for unknown contact forces, Lagrange multipliers, and associated boundary conditions to enforce filament inextensibility~\cite{moreau_asymptotic_2018,jabbarzadeh_numerical_2020}. The elastohydrodynamic PDE is recast into a simpler set of ordinary differential equations (ODEs) with trivial and intuitive matricial form, only depending on the filament parametrization, as we explore further in this paper. Moreau et al.~\cite{moreau_asymptotic_2018} highlighted the high efficiency of the formalism in 2D, and showed excellent agreement with classical formulations, even for relatively large coarse-graining segments. Following this, the coarse-graining method has gained momentum, and was adopted and generalised in several ways: (i) non-local hydrodynamics have been accounted for passive and active filaments in 2D, including wall interaction ~\cite{hall-mcnair_efficient_2019, walker_ishimoto_gadelha_gaffney_2019,walker_gaffney_2021}, (ii) 3D filament deformations were accounted for in ~\cite{el_alaoui-faris_optimal_2020,walker_efficient_2020}, though limited to a local hydrodynamic resistive theory (RFT), as in~\cite{moreau_asymptotic_2018},  (iii) it has been applied to study sperm  flagellum propulsion in 2D~\cite{neal_doing_2020}, (iv) expanded into a hybrid immersed boundary~\cite{ntetsika_numerical_2022}, as well as, stochastic method~\cite{ntetsika2021hybrid}, and (v) used for analytical and numerical study of the well-posedness of Newtonian and viscoelastic elastohydrodynamic propulsion~\cite{mori2022well,ohm2022well}.

Despite the above momentum in the literature, there is currently no coarse-graining formulation that allows simulation of collections of interacting elastohydrodynamic filaments with arbitrary solid structures in 3D. Here, we capitalise on the strengths of coarse-graining formulation and develop a fast simulation tool to resolve multiple interacting elastohydrodynamic filaments and solid body microhydrodynamics in 3D. Our generalised coarse-graining formulation is efficient, intuitive and easy-to-implement and customise. Our method allows for simulation of 3D elastic filaments with both flexural and torsional deformations, coupled hydrodynamically to free or fixed solid bodies and microstructures present in the environment (see Fig.~\ref{intro_fig}(c)). We use spheres as the hydrodynamic `building block' of filaments and solid structures for straightforward application of the method to novel and arbitrary three-dimensional geometries, though we note that any hydrodynamic theory may be employed instead. The method is validated against previous experiments and simulations in the literature.

We show that despite the substantial advantage of employing the coarse-graining formulation, runtimes can still be hindered by the choice of coordinate-system parametrization. We exploit the benefits of the exponential mapping of quaternions~\cite{boyle_integration_2017,rucker_integrating_2018} for high-precision tracking of basis-rotations in three-dimensions. This  allows fast and efficient computation, up to 150 times faster than a direct quaternion parametrisation. The exponential map also avoids undesired repeated change of basis used to circumvent coordinate-singularities introduced by Euler-angles \cite{nizette_towards_1999,el_alaoui-faris_optimal_2020,walker_efficient_2020}, critical whilst tracking multiple filaments simultaneously. Finally, we highlight the strengths of the new method in a series of non-trivial applications and report novel results. These include the beating of multi-flagellated swimmers,  sperm-egg elastohydrodynamic scattering, and particle transport by cilia arrays.

	\section{Methods }
 	We consider collections of 3D inextensible and unshearable elastic filaments undergoing deformations of bend and twist (Kirchoff rods) that are embedded in a viscous inertialess fluid, interacting hydrodynamically with each other and with solid bodies and micro-structures with arbitrary shapes, free or fixed, present in the micro-environment. We begin by outlining the coarse-graining formulation solving the dynamics of a single 3D elastohydrodynamic filament connected to a solid body, before generalising to an arbitrary number of filaments and solid bodies.
\begin{figure*}[t!]
    \includegraphics[width=\textwidth]{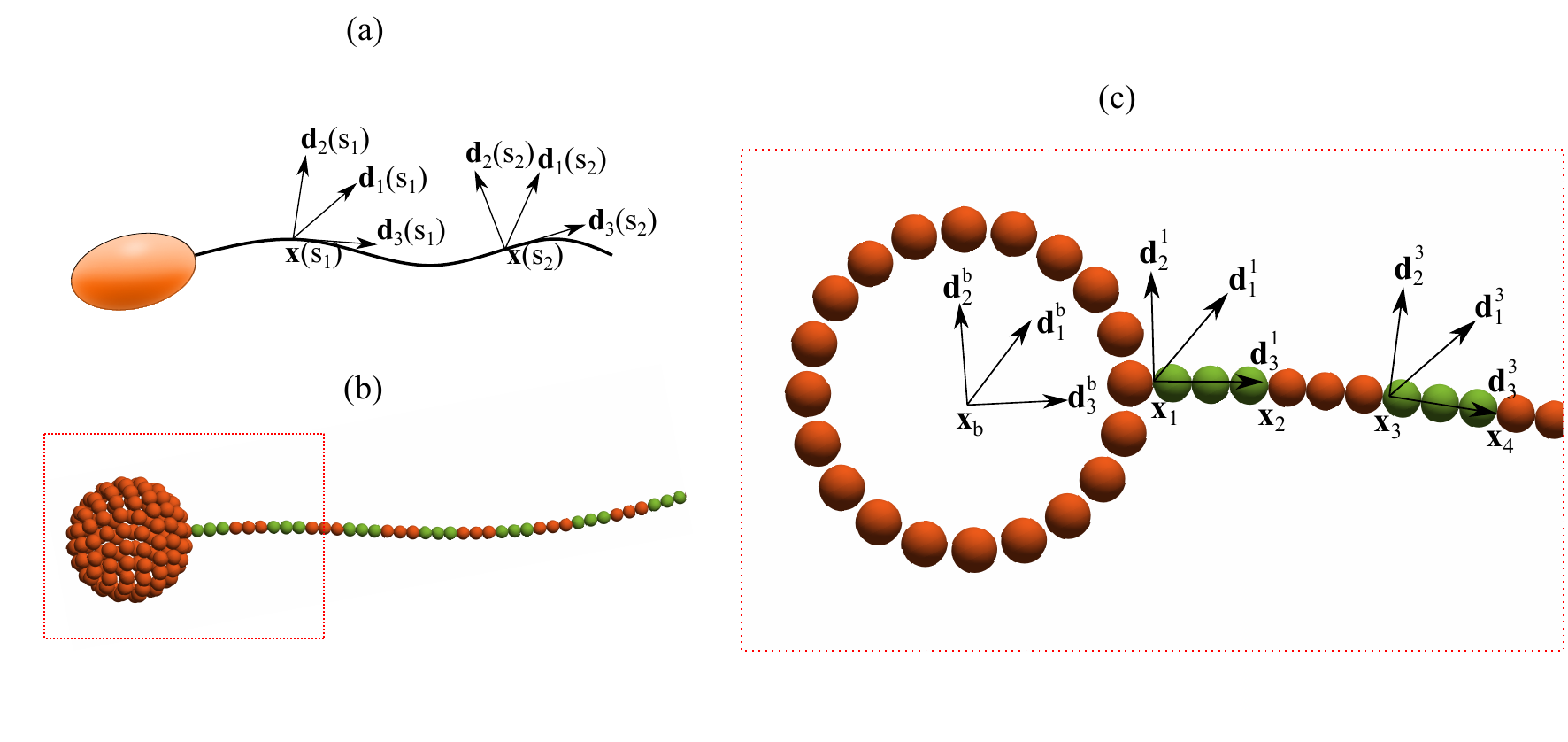}
    \caption{(a) An elastohydrodynamic structure comprised of solid body and filament. The filament is described by centreline $\mathbf{x}(s)$ with director basis $[\mathbf{d}_1(s), \mathbf{d}_2(s), \mathbf{d}_3(s)]$ shown at two arbitrary points $s = s_1$ and $s = s_2$. (b) The structure is made up of spheres. Body constructed using $N_{body} = 133$ spheres and filament using $N = 13$ rigid  segments with $n = 3$ spheres per segment. (c) Body position $\mathbf{x}_b$ and body director basis $[\mathbf{d}_1^b, \mathbf{d}_2^b, \mathbf{d}_3^b]$ labelled. Segment position $\mathbf{x}_i$ labelled for as well the the director basis for segments $1$ and $3$.}
    \label{method_fig}
\end{figure*}

	\subsection{The coarse-graining formalism in 3D}
		We consider first a single Kirchoff filament attached to an arbitrary solid body (Fig.~\ref{method_fig}(a)). Let $\mathbf{x}(s)$ be the position of the centreline of the filament, where $s$ is an arclength between $s = 0$ and $s = L$, where $L$ is the length of the filament. To describe deformations of the filament we will use the orthonormal local director basis $[\mathbf{d}_1(s), \mathbf{d}_2(s), \mathbf{d}_3(s)]$, defined at every point along the centreline. As such, $\mathbf{d}_3(s)$ is tangent to the curve $\mathbf{x}(s)$ and given by $
		\mathbf{d}_3(s) = \partial \mathbf{x}/\partial s$,
	and $\mathbf{d}_1(s)$ and $\mathbf{d}_2(s)$ point normal to the curve $\mathbf{x}(s)$. Derivatives of the director basis with respect to arclength are related to the twist vector $\boldsymbol{\kappa}(s)$ by
	\begin{equation}
		\frac{\partial \mathbf{d}_i}{\partial s} = \boldsymbol\kappa \times \mathbf{d}_i,
		\label{curv_dir}
	\end{equation}
	where $\boldsymbol{\kappa} = \kappa_1 \mathbf{d}_1 + \kappa_2 \mathbf{d}_2 + \kappa_3 \mathbf{d}_3$. The constituents of the twist vector $\kappa_1$, $\kappa_2$ and $\kappa_3$ are known as the filament curvatures \cite{nizette_towards_1999}. The third curvature in the $\mathbf{d}_3$ direction can also be referred to as the twist density. We direct the reader to Ref.~\cite{goriely_mathematics_2017} for a thorough discussion on the interplay between the three components of the twist vector and the differences between $\kappa_3$, torsion and twist.
	
	The balance of forces and torques along the filament at low Reynolds number reads \cite{antman_problems_2005}
	\begin{eqnarray}
		\mathbf{n}_s + \mathbf{f} &= 0 , \label{fba}\\
		\mathbf{m}_s + \mathbf{x}_s \times \mathbf{n} + \boldsymbol{\tau}  &=0 , \label{mba}
	\end{eqnarray}
	where $\mathbf{n}(s)$ and $\mathbf{m}(s)$ are the contact force and contact moment, $\mathbf{f}(s)$ and $\boldsymbol{\tau}(s)$ are the force and torque densities exerted on the filament by the fluid, and subscript $s$ denotes differentiation with respect to $s$. The elastic moment is known constitutively for a Kirchoff rod, whereas the contact force remains unknown. The coarse-graining formalism is derived from first principles by exploiting the integral form of the momentum balance above. This circumvents the need for solving a high-order PDE with a second-order boundary value problem using Lagrange multipliers \cite{moreau_asymptotic_2018}. For this, we discretise the body into $N_{body}$ spheres and the filament into $N$ rigid segments each made of $n$ spheres, as shown in Fig.~\ref{method_fig}(b) for $N_{body} = 133$, $N = 13$ and $n = 3$. Positions of spheres in the body are defined relative to the frame of reference that moves with the entire structure, $\mathbf{x}_b$, described by the local basis $[\mathbf{d}_1^b, \mathbf{d}_2^b, \mathbf{d}_3^b]$ (Fig.~\ref{method_fig}(c)). Sphere positions are given by $\mathbf{y}_i$ for $i = 1, 2, ..., M$, where $M = N_{body}+Nn$ is the total number of spheres in the system.  Body spheres are labelled by $i \leq N_{body}$ and filament spheres by $i > N_{body}$. Segment endpoints are located at $\mathbf{x}_i$ for $i = 1, 2, ..., N$, and $\mathbf{x}_1$ is defined relative to $\mathbf{x}_b$  in the local basis $[\mathbf{d}_1^b, \mathbf{d}_2^b, \mathbf{d}_3^b]$.

	Invoking {force- and torque-free conditions}, the total momentum balance of the elastohydrodynamic filament-body system in Eqs.~(\ref{fba},\ref{mba}) reads
	\begin{eqnarray}
		\sum_{i=1}^{N_{body} + Nn}\mathbf{F}_i &= 0,
		\label{total_force}\\
		\sum_{i=1}^{N_{body}+Nn} \left((\mathbf{y}_i - \mathbf{x}_b) \times \mathbf{F}_i + \mathbf{T}_i\right) &= 0,\label{total_torque}
	\end{eqnarray}
	where $\mathbf{F}_i$ and $\mathbf{T}_i$ are the total force and torque, respectively, exerted on sphere $i$ by the fluid. Integrating Eq.~\ref{mba} from $s = s_j$ to $s = L$ yields 
	\begin{equation}
		\sum_{i=N_{body}+(j-1)n+1}^{N_{body}+Nn} \left((\mathbf{y}_i - \mathbf{x}_j) \times \mathbf{F}_i + \mathbf{T}_i\right) = -\mathbf{m}_j,
		\label{sub_torque}
	\end{equation}
	for $j = 2, ..., N$, where $\mathbf{m}_j$ is the bending moment at $s_j$ defined constitutively by
	\begin{equation}
		\mathbf{m}_j = E_b \kappa_1^j \mathbf{d}_1^j + E_b \kappa_2^j \mathbf{d}_2^j  + E_t \kappa_3^j \mathbf{d}_3^j,
	\end{equation}
	where $E_b$ is the bending stiffness, $E_t$ is the torsional stiffness and $\kappa_i^j$ are the curvatures at $s_j$ for $i = 1,2,3$. The coarse-grained governing equations (\ref{total_force}, \ref{total_torque}, \ref{sub_torque}) can be written in a simpler and compact matrix form, given by
	 
\begin{widetext}
	\begin{equation}	
		\begin{bNiceArray}{wl{1cm}cccwr{1cm}wl{1cm}cccwr{1cm}}[first-col,code-for-first-col = \scriptscriptstyle ,]
		\text{Eq.~(\ref{total_force})}	&	\mathbf{I} 							 &  & \Cdots	&			& \mathbf{I}								& \mathbf{0}& & \Cdots&  & \mathbf{0} \\
		\text{Eq.~(\ref{total_torque})}   &	\left[\boldsymbol{\Delta}_1\right]_\times                &  & \Cdots	&		&	\left[\boldsymbol{\Delta}_M\right]_\times    	& \mathbf{I}& & \Cdots&& \mathbf{I} \\ 
		\text{Eq.~(\ref{sub_torque}) } j = 2	&	\mathbf{0}& \Cdots & \mathbf{0} & \Block{3-2}{\mathbf{A}} &	 & 				\mathbf{0} & \Cdots & \mathbf{0} & \Block{3-2}{\mathbf{B}} &  \\				
		  \Vdots	&	\Vdots	  &  \Ddots &\Vdots      &				 &	 & 				\Vdots	   &\Ddots &\Vdots      &               &  \\		
				\text{Eq.~(\ref{sub_torque}) } j=N	&	\mathbf{0}&\Cdots  & \mathbf{0} &                &   &              \mathbf{0} &\Cdots  &\mathbf{0}  &                &\\									
\CodeAfter
\tikz	
\draw [decorate, decoration = {brace}] ([xshift=-0mm,yshift = 2mm]1-1.north west) to node [auto = left] {$3M$} ([yshift = 2mm]1-5.north east);
\tikz	
\draw [decorate, decoration = {brace}] ([xshift=-0mm,yshift = 2mm]1-6.north west) to node [auto = left] {$3M$} ([yshift = 2mm]1-10.north east);
		\end{bNiceArray}
	\begin{bmatrix}
	\mathbf{F}_1 \\ \vdots	\\ \mathbf{F}_M \\ \mathbf{T}_1 \\ \vdots \\ \mathbf{T}_M\end{bmatrix} 
	= -\begin{bmatrix}
		\mathbf{0} \\ \mathbf{0} \\ \mathbf{m}_2 \\ \vdots \\ \mathbf{m}_N
	\end{bmatrix},
\label{Mf_fil_body}
	\end{equation}
\end{widetext}
	where we have used the matrix form of the cross product, denoted by $ [\mathbf{a}]_\times$ {(see Appendix \ref{appen_cross})}, and defined $\boldsymbol{\Delta}_i = \mathbf{y}_i - \mathbf{x}_b$. The blocks of zeros multiply forces and torques experienced by spheres in the body, as they do not contribute to the bending moment relations on the filament. The matrices $\mathbf{A}$ and $\mathbf{B}$ multiply forces and torques experienced by the spheres in the filament, and are size $3(N-1)\times3Nn$. The matrix $\mathbf{A}$ is comprised of $(N-1)Nn$ sub-matrices given by $A_{i,j} = [\mathbf{y}_{N_{body}+j} - \mathbf{x}_{i+1}]_\times, \text{ if } j > in$, and zero otherwise. The matrix $\mathbf{B}$ has the same structure as $\mathbf{A}$ with sub-matrices given by $B_{i,j} = \mathbf{I}, \text{ if } j > in$, and zero otherwise. 

	The coarse-grained matrix formulation in Eq.~\ref{Mf_fil_body} encodes the force and torque balance for a single filament attached to an arbitrary solid body (Fig.~\ref{method_fig}(b)). By setting $N_{body}=0$ or $N = 0$, we model a free filament and a free solid body, respectively, further highlighting the simplicity of the formulation. The general form of the coarse-grained formulation in Eq.~\ref{Mf_fil_body} for an arbitrary number of solid bodies each with an arbitrary number of attached filaments is simply given by
	\begin{equation}
		\boldsymbol{\mathcal{M}}_F \begin{bmatrix}
			\boldsymbol{\mathcal{F}} \\ \boldsymbol{\mathcal{T}}
		\end{bmatrix} = \boldsymbol{\mathcal{K}},
	\label{general_Mf}
	\end{equation}	
	where $\boldsymbol{\mathcal{M}}_F$ is analogous to the matrix in Eq.~\ref{Mf_fil_body}, $\boldsymbol{\mathcal{F}}$ and $\boldsymbol{\mathcal{T}}$ contain the forces and torques, respectively, exerted on all spheres by the fluid. $\boldsymbol{\mathcal{K}}$ is analogous to the right-hand-side in Eq.~\ref{Mf_fil_body} and defines the constitutive relation for each 3D Kirchoff  filament in the system. In the Appendix ~\ref{appendix:two_free_filaments}, we provide the explicit form of Eq.~\ref{general_Mf} and include a practical example of the coarse-grained matrix in the case of two free filaments. It is worth noting that the coarse-graining formulation in Eq.~\ref{general_Mf} is general and does not depend on the specific choice of hydrodynamic coupling. For this reason, Eq.~\ref{general_Mf} can be used in conjunction with a variety of computational and analytic methods for solving the fluid mechanics around slender bodies. In this paper, we exploit the mathematical capabilities of the Rotne-Prager-Yamakawa (RPY) non-local hydrodynamic approximation  \cite{zuk_rotneprageryamakawa_2014} for low Reynolds number fluids.  

\subsection{Non-local hydrodynamic coupling and dimensionality reduction}
 We approximate the non-local hydrodynamic forces and torques on each interacting sphere using the Rotne-Prager-Yamakawa (RPY) tensor  \cite{zuk_rotneprageryamakawa_2014}. This relates linear velocities $\mathbf{v}_i$ and angular velocities $\boldsymbol{\omega}_i$  with forces and torques acting on each sphere via the non-local hydrodynamic mobility matrix $\boldsymbol{\mathcal{M}}_H$, giving
\begin{equation}
	\begin{bmatrix}
		\boldsymbol{\mathcal{V}} \\ \boldsymbol{{\Omega}}
	\end{bmatrix}= \boldsymbol{\mathcal{M}}_H	\begin{bmatrix}
		\boldsymbol{\mathcal{F}} \\ \boldsymbol{\mathcal{T}}
	\end{bmatrix},
\end{equation}
where $\boldsymbol{\mathcal{V}}$ and $\boldsymbol{{\Omega}}$ contain the velocities  $\mathbf{v}_i$ and angular velocities $\boldsymbol{\omega}_i$ of all spheres. The mobility tensor encodes the translational and rotational movement of each sphere, as well as the non-local coupling between spheres, depending on the sphere radius $a_i$, fluid viscosity $\eta$ and mutual distance,  please see Appendix \ref{Hydro-appen} for details. 

Exploiting the invertibility of the non-local hydrodynamic mobility tensor above, the generalised coarse-graining formulation (Eq.~\ref{general_Mf}) for the \emph{non-local} elastohydrodynamic coupling simply reads  
\begin{equation}
	\boldsymbol{\mathcal{M}}_F     \boldsymbol{\mathcal{M}}_H^{-1} \begin{bmatrix}
		\boldsymbol{\mathcal{V}} \\
		\boldsymbol{{\Omega}}
	\end{bmatrix} = \boldsymbol{\mathcal{K}}.
	\label{MH_sys}
\end{equation}
However, the governing Eq.~\ref{MH_sys} lists, unnecessarily, all linear $\mathbf{v}_i$ and angular $\boldsymbol{\omega}_i$ velocities for each filament's sphere as unknowns. Within the same body-filament structure, translation and rotations of each is sphere are coupled, as they are part of the same assembly. Hence,  linear velocities can be written in terms of the angular velocities to reduce the dimensionality of a given body-filament structure (Fig.~\ref{method_fig}), see Appendix~\ref{reduction_dimension} for details. As a result, $\mathbf{v}_i$ and $\boldsymbol{\omega}_i$ are a function of the velocity of the centre of mass of the body $\dot{\mathbf{x}}_b$, its angular velocity $\boldsymbol{\omega}_b$, and the angular velocity of each segment $\boldsymbol{\omega}_i^{seg}$, part of the same body-filament structure. By rigidly attaching the first segment to the body (Fig.~\ref{method_fig}), we set $\boldsymbol{\omega}_1 = \boldsymbol{\omega}_b$. Altogether, this reduces the dimensionality of the system from $6(N_{body}+Nn)$ to $3+3N$ in Eq. (\ref{MH_sys}). The dimensionality reduction is achieved via
\begin{equation}
	\begin{bmatrix}
		\boldsymbol{\mathcal{V}} \\ \boldsymbol{\Omega}
	\end{bmatrix} = \boldsymbol{\mathcal{Q}} \boldsymbol{\mathcal{W}},
	\label{Q_1}
\end{equation}
where $\boldsymbol{\mathcal{W}}$ is the reduced velocity system for each sphere in terms of  ${\mathbf{x}}_b, {\boldsymbol{\omega}_b},  {\boldsymbol{\omega}_i^{seg}}$, for $i > 1$. Substituting Eq.~(\ref{Q_1}) into~(\ref{MH_sys}) gives the full non-local, coarse-graining elastohydodynamic system in 3D 
\begin{equation}
		\boldsymbol{\mathcal{M}}_F     \boldsymbol{\mathcal{M}}_H^{-1} \boldsymbol{\mathcal{Q}} \boldsymbol{\mathcal{W}} = \boldsymbol{\mathcal{K}},
		\label{system_after_Q}
\end{equation}
where $\boldsymbol{\mathcal{W}}$ contains the reduced unknown velocities of each interacting unit of the elastohydrodynamic system. Knowledge of $\boldsymbol{\mathcal{W}}$ provide linear and angular velocities of each sphere, but without the filament shape information. The filament shape, position and orientation can be found by numerically integrating these velocities relative to the laboratory fixed frame of reference, for example, using the rate of change of the directors base
$\mathrm{d} \mathbf{d}_i/{\mathrm{d} t} = \boldsymbol{\omega} \times \mathbf{d}_i $. The latter however introduces severe accuracy issues due to preserving orthogonality and unit length of the directors \cite{boyle_integration_2017}. We circumvent these difficulties in the next section by exploiting the exponential mapping of quaternions to accurately track rotations of the basis vectors in 3D.

\subsection{Tracking director rotations in 3D using exponential mapping of quaternions}
Euler angles are commonly used to track 3D rotations of vector basis as function of three parameters. The challenge of any three-parameter parametrisation is the existence of coordinate-singularities. With Euler angles this corresponds to performing two successive rotations about the same axis, effectively losing a degree of freedom, known as gimbal lock \cite{grassia_practical_1998}. Coordinate-singularities can be circumvented by continuously changing the laboratory frame of reference, though this method is inappropriate for tracking multiple interacting filaments at once as we require here. Quaternions are canonically used to avoid coordinate-singularities by introducing four-parameter rotations, although, this can increase considerably the computational time, as we detail in the results section. Instead, we focus our study on the exponential mapping of quaternions to track rotations in 3D (referred here as exponential mapping). The exponential map effectively reparametrises quaternions into a three-parameter rotation scheme, thus re-introducing singularities. However, unlike Euler angles, the coordinate-singularities of the exponential map are trivial to avoid, requiring only a re-scaling rather than redefining entirely the laboratory frame of reference \cite{grassia_practical_1998}. This enables faster and more efficient simulations of multiple interacting filaments and solid bodies via coarse-graining formulation introduced above. 
\begin{figure*}[t!]
    \includegraphics[width=\textwidth]{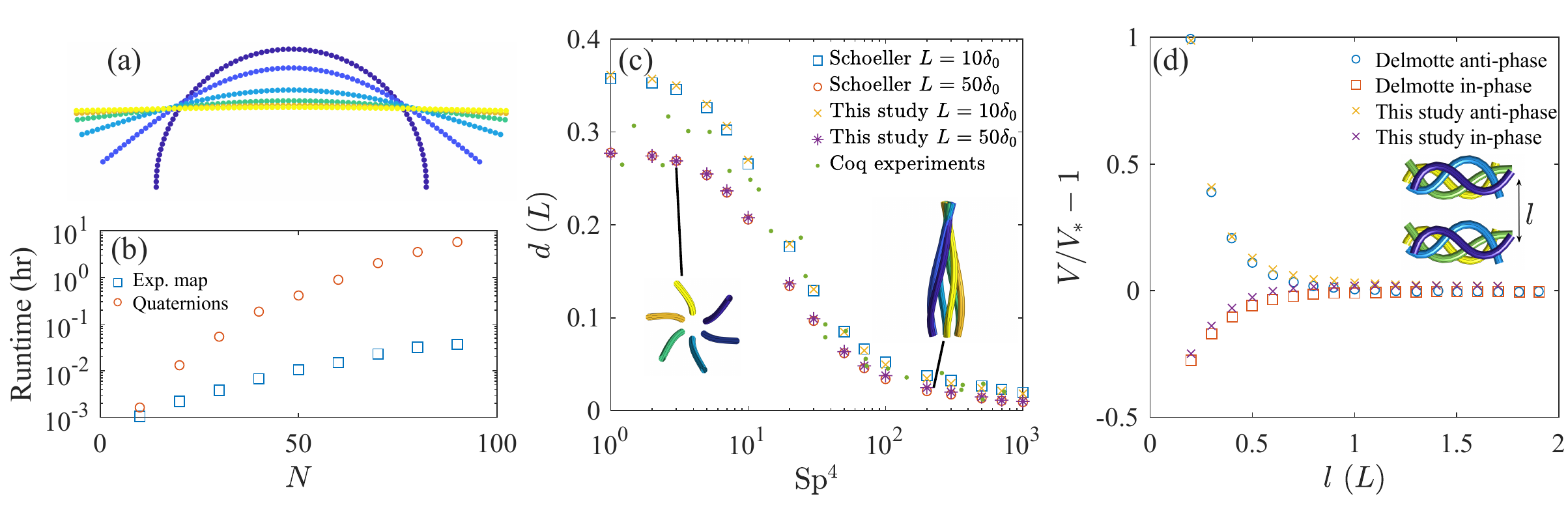}
    \caption{(a) Relaxation of filament with $N = 40$, $n = 2$ and $\mathcal{S} = 3$. Initial condition dark blue and shown at subsequent times $t = 0.5, 1, 1.5, 2, 2.5$, ending in a straight configuration. (b) Runtime in hours to solve for relaxation of single filament using quaternion and exponential map parametrisation, for various number of segments $N$. Number of spheres per segment is $n = 2$ for all cases. (c) Distance that the end tip of helically actuated filament makes with the axis of rotation, averaged over time during steady state. Sperm number $\mathrm{Sp}^4 = ((4\pi)(\log{L/a}+0.5))\mathcal{S}^4$. Results agree well with simulations from Schoeller et al. \cite{schoeller_methods_2021} and experiments from Coq. et al \cite{coq_helical_2009}. (d) Swimming speed of two swimmer system normalised by swimming speed of a single swimmer. Inset shows in-phase swimmers at several times over one period, with the initial distance between swimmers $l$ labelled. Results show good agreement with data from Delmotte et al. \cite{delmotte_general_2015}.}
    \label{validations}
\end{figure*}

Next, we will briefly outline quaternions and exponential mapping, and their coupling with coarse-graining formulation. Quaternions are an extension to the complex plane into a complex four-dimensional space. A quaternion is given by
$\mathbf{q} = q_0 + q_1 i + q_2 j + q_2 k = \begin{bmatrix}
		q_0 & q_1 & q_2 & q_3
	\end{bmatrix}^\top$, where $i^2 = j^2 = k^2 = ijk = -1$. Unit quaternions can be used to rotate a vector basis $\mathbf{v}$ in space via $\mathbf{v}\prime = \mathbf{q}\mathbf{v}\mathbf{q}^{-1}$. The axis of rotation is given by the imaginary part of $\mathbf{q}$ and the angle of rotation by the real part of $\mathbf{q}$. Specifically,
\begin{equation}
	\mathbf{q} = \begin{bmatrix}
		\cos({\theta/2}) \\ \sin{(\theta/2)}\hat{\mathbf{n}}
	\end{bmatrix}
	\label{theta_ax_q}
\end{equation}
gives the quaternion corresponding to a rotation about the axis $\hat{\mathbf{n}}$ by $\theta$. The angular velocity of a set of basis vectors is related to the quaternion by
	\begin{equation}
	\dot{\mathbf{q}} = \frac{1}{2}\begin{bmatrix}
		-q_1 & -q_2 & -q_3 \\
		q_0 & q_3 & -q_2 \\
		-q_3 & q_0 & q_1 \\
		q_2 & -q_1 & q_0
	\end{bmatrix}
	\boldsymbol \omega,
	\label{qi_omega}
\end{equation}
providing a simple framework for numerical integration of quaternions, though this increases the dimensionality of the system. Generalising above to account for the multiple angular velocities of the system $\boldsymbol{\mathcal{W}}$, the state shape-vector  $\boldsymbol{\mathcal{X}}_{quat}$ evolve in time according to
\begin{equation}
	\dot{\boldsymbol{\mathcal{X}}}_{quat} = \boldsymbol{\mathcal{C}}\boldsymbol{\mathcal{W}},
		\label{quat}
\end{equation}
and is solved together with governing Eq.~(\ref{system_after_Q}), where the matrix $\boldsymbol{\mathcal{C}}$ is given in Appendix.~\ref{appendix:quaternions}.

The exponential mapping, on the other hand, re-parameterises quaternions back to a three-parameter parametrization \cite{boyle_integration_2017}. We define a pure vector known as the generator $\mathbf{r}$, which is related to a quaternion via the exponential map, $\mathbf{q} = \exp({\mathbf{r}}) = \begin{bmatrix}
		\cos{|\mathbf{r}|} & \sinc{|\mathbf{r}|}\mathbf{r}^\top
	\end{bmatrix}^\top$, where $\sinc{x} = \sin{x}/x$. The angular velocity of a set of basis vectors defined by generator $\mathbf{r}$ can be written conveniently in matrix form \cite{boyle_angular_2013}, 
\begin{equation}
 \boldsymbol{\omega} = 2\mathbf{D}\dot{\mathbf{r}},
 \label{2Dr}
\end{equation}
where the matrix $\mathbf{D}$ is given by
\begin{equation}
	\begin{aligned}
		\mathbf{D}=&\left\{\left[\begin{array}{lll}
			1 & 0 & 0 \\
			0 & 1 & 0 \\
			0 & 0 & 1
		\end{array}\right]+\frac{\sin ^{2} |\mathbf{r}|}{|\mathbf{r}|^{2}}\left[\begin{array}{ccc}
			0 & -r_3 & r_2 \\
			r_3 & 0 & -r_1 \\
			-r_2 & r_1 & 0
		\end{array}\right]\right.\\
		&\left.-\frac{|\mathbf{r}|-\sin |\mathbf{r}| \cos |\mathbf{r}|}{|\mathbf{r}|^{3}}\left[\begin{array}{ccc}
			r_2^{2}+r_3^{2} & -r_1 r_2 & -r_1 r_3 \\
			-r_1 r_2 & r_1^{2}+r_3^{2} & -r_2 r_3 \\
			-r_1 r_3 & -r_2 r_3 & r_1^{2}+r_2^{2}
		\end{array}\right]\right\}
	\end{aligned}.
	\label{eqn:matrixA}
\end{equation}
The determinant $|\mathbf{D}| = \sinc^2 |\mathbf{r}|$ goes to zero as $|\mathbf{r}| \to n\pi$, for integer $n \geq 1$. As such, a rescaling is necessary whenever the magnitude of a generator approaches $n\pi$ to avoid coordinate-singularity \cite{boyle_integration_2017}. This can be achieved by rescaling  
	\begin{equation}
	\mathbf{r} \to \mathbf{r} - \pi \frac{\mathbf{r}}{|\mathbf{r}|},
\end{equation}
whenever $|\mathbf{r}| \geq \pi/2$. This rescaling does not change the rotation performed by the generator $\mathbf{r}$ but keeps it far away from any coordinate-singularity. Eq.~(\ref{2Dr}) can be generalised to account for the angular velocities of each sphere in the system $\boldsymbol{\mathcal{W}}$ so that the state shape-vector  $\boldsymbol{\mathcal{X}}_{gen}$ can be evolved in time using 
\begin{equation}
	\boldsymbol{\mathcal{W}} = \boldsymbol{\mathcal{D}}\dot{\boldsymbol{\mathcal{X}}}_{gen},
\end{equation}
which can be substituted, directly, into the governing system of equations~(\ref{system_after_Q}). This provides a convenient matrix form to evolve in time the non-local, multi-filament-body elastohystodynamic ODE system 
\begin{equation}
	\boldsymbol{\mathcal{M}}_F\boldsymbol{\mathcal{M}}_H^{-1}\boldsymbol{\mathcal{Q}}\boldsymbol{\mathcal{D}}\dot{\boldsymbol{\mathcal{X}}}_{gen} = \boldsymbol{\mathcal{K}}.
	\label{sys_expmap}
\end{equation}
After prescribing an initial state vector and boundary conditions, the governing system of equations, Eqs.~(\ref{system_after_Q},\ref{quat}) for quaternions, and Eq.~(\ref{sys_expmap}) for exponential mapping, can be solved directly for the time evolution of the system's configuration state. We approximate curvatures in $\boldsymbol{\mathcal{K}}$ using finite differences of the director basis (using Eq.~\ref{curv_dir}). We use MATLAB solver \textit{ode15s} to handle numerical integration, but note that any ODE solver can be used, highlighting the simplicity of the matrix representation. A Matlab code is provided to serve as a basis for rapid customization and further generalizations.

\section{Exponential mapping of quaternions and comparison with PDE formulations}
We proceed with a comparison between  the computational efficiency of quaternions and exponential mapping within the coarse-graining framework. We also compare our simulations with previous methods with similar levels of elastohydrodynamic accuracy 
\cite{schoeller_methods_2021,delmotte_general_2015}, which have been equally validated against experiments, though employing distinct fluid-structure interaction approaches. Finally, in the next section, we provide exemplars of the capability of the method to a series of complex elastohydrodynamic systems in three-dimensions that are straightforward to implement using the CG formulation. 

We non-dimensionalise the system using length-scale $L$, timescale $T$ given either by the characteristic frequency $\omega$ of the system or the relaxation time of the fibre, and force density $E_b / L^3$.  For simplicity, we set $E_b = E_t$.
Finally, we consider any filament attachment to the body as a rigid connection, i.e. the filaments are clamped at body.
The dimensionless system of equations is thus multiplied by the non-dimensional stiffness parameter	$\mathcal{S}^4 = L^4 \frac{\eta \omega}{E_b}$, where $\eta$ is the fluid viscosity. The stiffness parameter is closely related to parameters, such as, the elastohydrodynamic number $\mathrm{E_h} = 8\pi\mathcal{S}^4$, and the sperm number $\mathrm{Sp}^4=(\frac{\xi_\perp}{\eta} \mathcal{S}^4)$, where $\xi_\perp$ denotes the perpendicular resistive force drag coefficient from RFT used elsewhere  \cite{moreau_asymptotic_2018,walker_efficient_2020}, and characterises the ratio of viscous and elastic forces in the system.

We first compare the computational performance between quaternion and exponential mapping for the simple case of a filament free from external forces and torques, initially bent and  undergoing relaxation dynamics, by recording the computational time taken to solve for the dynamics. Given no body exists, $\mathbf{x}_b = \mathbf{x}_1$, $\mathbf{q}_b = \mathbf{q}_1$ and $\mathbf{r}_b = \mathbf{r}_1$. Initial quaternions are given by $\mathbf{q}_i = \begin{bmatrix}\cos({\theta_i/2}) & \sin{(\theta_i/2)}\hat{\mathbf{n}}^\top \end{bmatrix}^\top$ and generators by $\mathbf{r}_i = \theta_i/2\hat{\mathbf{n}}$,
where $\theta_i = (i-1)\pi/(N-1)$ and $\hat{\mathbf{n}}= \begin{bmatrix}
	0 & 1 & 0
\end{bmatrix}$, for $i = 1, ..., N$. This corresponds to a filament bent into a semi-circle, as shown in Fig.~\ref{validations}(a) for $N=40$, $n = 2$. The initial shape unbends towards the unstressed straight configuration, with the centre of mass position conserved to within $2\times10^{-2}$ $(L)$. We set $n = 2$ and vary the number of segments $N$. Figure.~\ref{validations}(b) shows the runtime in hours to solve from $t = 0$ to $t = 20$, for various values of $N$ for both the quaternion and exponential mapping, with the stiffness parameter $\mathcal{S} = 3$. For $N = 10$ segments, runtimes are comparable at 3.8 s and 5.6 s for the exponential map and quaternions, respectively. As $N$ increases the exponential map results in significantly reduced runtimes. For example, at $N = 30$, runtimes are 13.7 s and 3 min for the exponential map and quaternions, respectively. This difference in runtime increases exponentially, and for $N = 90$ segments the exponential map parametrisation runs over 150 times faster. This increase in runtime is true across all ranges of $\mathcal{S}$. Hereafter, we proceed our numerical study employing exponential mapping for more efficient computation.

In order to further test that deformations and displacements of the filament are being simulated accurately, we compare our results with experiments and simulations carried out by previous works. Schoeller et al. \cite{schoeller_methods_2021} simulate a single helically actuated filament, where one end of the filament is held a constant distance $\delta_0$ and angle $\alpha_0$ from an axis of rotation. The base of the filament is rotated at an angular velocity $\omega$ about the axis of rotation.  We implement this using a filament with $N = 20$ and $n = 1$. Initial generators are all set to $\mathbf{r}_i = \alpha_0/2\begin{bmatrix}0 & 1& 0\end{bmatrix}^\top$. To enforce the motion of the base of the filament, we replace the force-free condition with the kinematic constraint $\dot{\mathbf{x}}_{1} = -\delta_0\begin{bmatrix}
		\sin{t} & \cos{t} & 0
	\end{bmatrix}^\top$, and the torque-free condition with the kinematic constraint,
$
	\dot{\mathbf{r}}_{1} = \frac{\alpha_0}{2}\begin{bmatrix}
		\cos{t} & -\sin{t} & 0 
	\end{bmatrix}^\top.$
We solve the system between $t = 0$ and $t = 10000$, ensuring the system reaches a steady state. Once a steady periodic state is achieved, we measure the distance that the tip of the filament makes with the axis of rotation, averaged over one period. Figure.~\ref{validations}(c) shows the tip distance compared with results from Ref.~\cite{schoeller_methods_2021} for $\delta_0 = 0.1 (L)$ and $\delta_0 = 0.02 (L)$ with $\alpha_0 = 0.2618$. Results agree excellently with both Ref.~\cite{schoeller_methods_2021} and experimental data from Ref.~\cite{coq_helical_2009}. It is worth noting that  Ref.~\cite{schoeller_methods_2021} has also been validated and contrasted with experiments independently.  The agreement with canonical elastohydrodynamic methods in 3D, such as in  \cite{schoeller_methods_2021}, {\bd }using the moment balance PDEs directly, together with Lagrange multipliers to enforce the inextensibility constraint (not required here), indicates equivalence between canonical PDE methods and the simpler coarse-grained method employed here in 3D, as also demonstrated in 2D by~\cite{moreau_asymptotic_2018}. 

We conduct one further test that places significance on the non-local hydrodynamics between spheres in the system. Following Delmotte et al.~\cite{delmotte_general_2015}, we simulate the case of a single swimmer and then two identical swimmers coupled via non-local hydrodynamics. We prescribe a time dependent intrinsic curvature to a filament made of $N = 20$ segments with $n = 1$ spheres per segment so that our hydrodynamic description matches the beads model employed in Ref.~\cite{delmotte_general_2015}. We use the curvature $\boldsymbol{\kappa}^0(s,t) = -K_0 \sin{(ks-\omega t)}\mathbf{d}_1(s,t)$, with $K_0 =8.25$, if $s \leq 0.5L$, and $K_0 =16.5(L-s)/L$, if $s > 0.5L$, and a fixed stiffness parameter $\mathcal{S} = 22.6^{1/4}$. Numerical parameter values are chosen to match Ref.~\cite{delmotte_general_2015}, which itself validated this swimming case with experiments \cite{2013APS..MARY39002B} and simulations \cite{majmudar_experiments_2012}. For the case of a single swimmer, the curvature wave causes planar beating that propels the filament forwards, characterized by a distance travelled per stroke of $V = 0.0671 (L/T)$. This swimming speed compares well with $V=0.066 (L/T)$ from simulations in Ref.~\cite{delmotte_general_2015} and $V = 0.07 (L/T)$ from experiments in Ref.~\cite{2013APS..MARY39002B}. 

We also measure the swimming speed of two active filaments swimming together, Fig~\ref{validations}(d), both with the same time-dependent curvature, as the single swimmer above, but set to either in-phase or anti-phase (inset of Fig~\ref{validations}(d) shows in-phase swimmers). The filaments are set an initial distance $l$ apart and their initial configuration is taken from the steady-state of the single swimmer case. The normalised swimming speed of the two-swimmer system is given in Fig.~\ref{validations}(d), and shows a reduction in swimming speed for in-phase swimmers as they approach each other, and an increase in swimming speed for anti-phase swimmers as they approach each other. Figure.~\ref{validations}(d) shows excellent agreement with Ref.~\cite{delmotte_general_2015} for in-phase and anti-phase swimmers. This is despite major differences in both modelling and computational formulations. In particular, Ref.~\cite{delmotte_general_2015} employs a novel gears model which solves the moment balance PDE with contact-contact forces between interlocked spheres making up the elastic filament which rotate relative to each other similarly to a gear, in such a way that the unknown contact forces play the role of Lagrange multipliers in the system. For anti-phase swimmers at small $l$, we used a modified version of the RPY tensors allowing for overlapping spheres given in Ref.~\cite{zuk_rotneprageryamakawa_2014}, as the beats of the two swimmers begin to overlap slightly \cite{delmotte_general_2015}.

\section{3D non-local Elastohydrodynamic applications}

\begin{figure*}
    \centering
    \includegraphics[width=\textwidth]{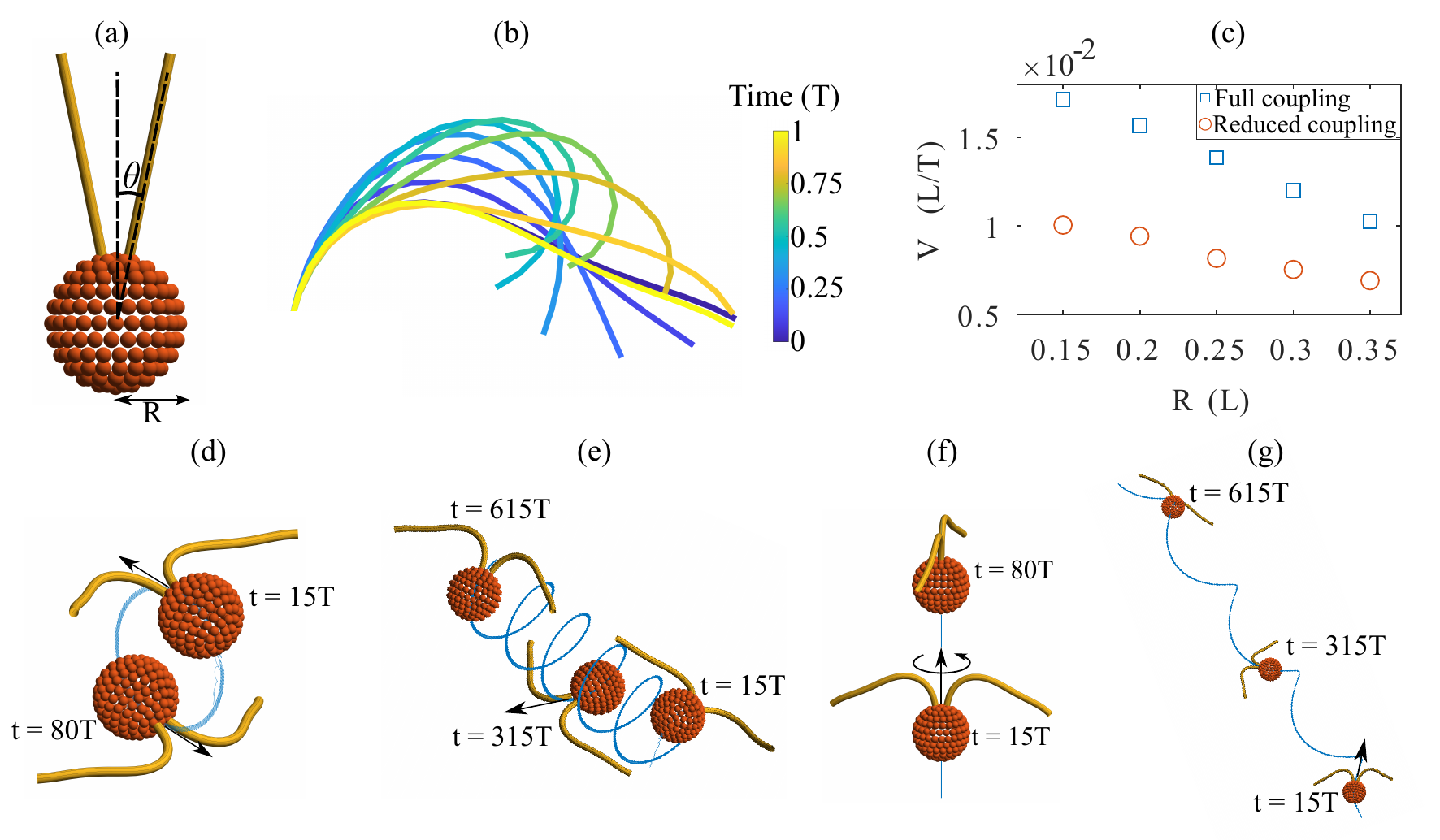}
    \caption{(a) Model \textit{Chlamydomonas} setup. Angle $2\theta$ between two flagella attached to body of radius $R$. Here, $2\theta=23$ degrees and $R = 0.35 (L)$. (b) Periodic bending wave given by time-dependent intrinsic curvature $\boldsymbol{\kappa} = \pm(4+4\sin(2\pi s-t)\mathbf{d}_2$ shown over one period $T$. Induces straight swimming of \textit{Chlamydomonas} indicated by arrow. The $\mathbf{d}_2$ director basis along the entire length of both filaments points directly into the page, meaning the bending wave produced by this time-dependent curvature is perfectly in the plane of the page. (c) Swimming speed of \textit{Chlamydomonas} dependent upon body radius and non-local hydrodynamic coupling. Reduced coupling indicates that terms in the RPY tensors coupling spheres in separate parts of the \textit{Chlamydomonas} are set to zero (hydrodynamic coupling between flagellum-flagellum and flagellum-body are turned off). (d) Adding a constant out of plane curvature of magnitude $\kappa_1$ in the $\mathbf{d}_1$ direction causes out of plane swimming. Trajectory of \textit{Chlamydomonas} when both flagella have equal out of plane intrinsic curvature $\kappa_1 = 0.5$. (e) Adding asymmetry between flagella turns the circular trajectory into a helical trajectory. Here, the flagella have intrinsic out of plane curvature $\kappa_1 = 0.5\pm0.3$. (f) Trajectory of \textit{Chlamydomonas} when both flagella have equal and opposite out of plane intrinsic curvature $\kappa_1 = \pm 0.5$. (g) Asymmetric opposite sign out of plane curvatures $\kappa_1 = \pm 0.5 + 0.3$ turns straight swimming with rolling into a helical trajectory, with rolling still present.}
    \label{fig:chlam_fig}
\end{figure*}
\subsection{Bi-flagellate \textit{Chlamydomonas} elastohydrodynamic-swimming in 3D}
\textit{Chlamydomonas} is a widespread model organism whose flagella are structured similarly to those found in mammals, making it an important microorganism in experiments and modelling of flagella motion \cite{sasso_molecular_2018}. However, modelling \textit{Chlamydomonas} is not an easy task due to the multiple elastic and solid interacting parts. It is composed of a solid body and two elastic tail-like appendages coupled by contact-contact and hydrodynamic interactions, whilst the whole system is free from external forces and torques during free-swimming motion. As such \textit{Chlamydomonas} modelling is often abstracted with minimal models \cite{friedrich_flagellar_2012,bennett_emergent_2013,cortese_control_2021}. Fauci et al. \cite{fauci_computational_1993,fauci_computational_1996} modelled \textit{Chlamydomonas} using the immersed boundary method, but since these studies we are unaware of any simulation work of \textit{Chlamydomonas} that includes full elastohydrodynamic interactions. Here, we demonstrate that simulating this model organism is straightforward with the coarse-graining formalism presented here.

\begin{figure*}[t!]
	\includegraphics[width = \textwidth]{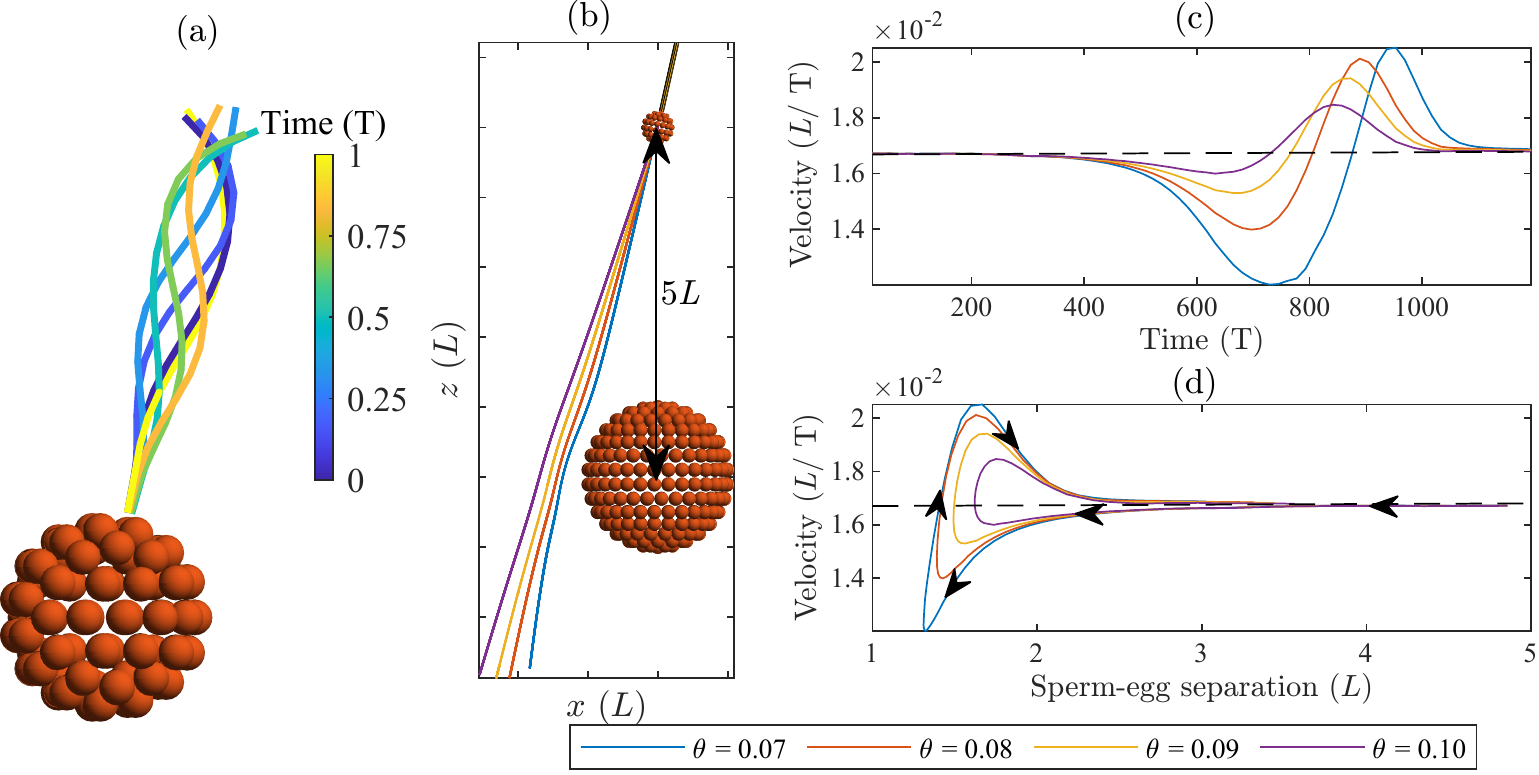}
	\caption{(a) Waveform of model sperm shown over one beat period T. (b) Trajectory of sperm past egg with different initial condition for $\theta$, the angle between the flagellum and $\hat{{z}}$. (c) Swimming speed dependent on time. Sperm slows down as it gets closer to the egg, and then begins to speed up as the sperm swims away from the egg. (d) Swimming speed dependent on sperm-egg separation, shown from $t = 0 (T)$ to $t = 1200 (T)$. Arrows indicate the flow of time. Depending on $\theta$, the sperm reaches a minimum separation from the egg before swimming away and increasing in speed.}
	\label{fig:model_sperm}
\end{figure*}

The coarse-grained \textit{Chlamydomonas} swimmer is made of two active elastic filaments with an angle $2\theta$ between them, attached to single spherical body of radius $R$, as depicted in Fig.~\ref{fig:chlam_fig}(a). The body is made out of $N_{body} = 184$ spheres and the filaments contain $N = 15$ segments with $n = 1$ sphere per segment. With $\mathcal{S} = 3$, we induce straight swimming by prescribing an intrinsic time-dependent curvature $\boldsymbol \kappa = \pm4[1+\sin(2\pi s - t)]\mathbf{d}_2(s)$ in each filament. The constant term in the prescribed curvature induces a non-zero average curvature along the flagella (see Fig.~\ref{fig:chlam_fig}(b)), allowing the flagella to progressively `pull' the spherical body over time, known as a puller swimmer. We vary the body radius (keeping $N_{body}$ = 184) and record the swimming speed of the \textit{Chlamydomonas} swimmer. Fig.~\ref{fig:chlam_fig}(c) shows the swmimming speed of each \textit{Chlamydomonas} as body size is increased with full non-local hydrodynamic coupling and reduced coupling (where hydrodynamic coupling between flagellum-flagellum and flagellum-body are turned off). Increasing the body size linearly decreases the swimming speed of \textit{Chlamydomonas}, as expected from the linearity of Stokes flow. Reduced hydrodynamic coupling results in a slower swimming speed (up to 60\% slower for the smallest body radius used here), highlighting the importance of non-local coupling for \textit{Chlamydomonas} swimming (not possible with minimal models), and how this microorganism may exploit this effect to achieve higher swimming speeds. The anti-phase~\textit{Chlamydomonas} beating is thus critical to unlock higher propulsion, similarly to hydrodynamic coupled single-swimmers in Fig.~\ref{validations}(d), that achieve higher speeds when moving in anti-phase near to each other. Interestingly, the difference in swimming speeds between the full and reduced hydrodynamic coupling decrease with the body size in Figure.~\ref{fig:chlam_fig}(c), indicating that Stokes law for the spherical body could provide a good approximation when the body is large (or the flagella is small). This because the non-local hydrodynamic effect becomes increasingly small, when compared with the zeroth order contribution from Stokes law, as the body size decreases. 

Motivated by recent empirical studies observing three-dimensional beating  of \textit{Chlamydomonas} flagella \cite{cortese_control_2021,wilson_chiral_2022}, we show here that it is straightforward to implement 3D beating of \textit{Chlamydomonas}  using the CG framework. For the simplicity of this demonstration, we introduce a constant out-of-plane curvature $\kappa_1$ in the $\mathbf{d}_1$ direction of the two flagella, so that  the intrinsic time-dependent curvature is $\boldsymbol \kappa = \kappa_1\mathbf{d}_1(s)\pm4[1+\sin(2\pi s - t)]\mathbf{d}_2(s)$. In the symmetric case, when $\kappa_1$ is the same for the two flagella, the  out-of-plane beating causes the swimming path of \textit{Chlamydomonas} to be no longer straight.
Instead, the \textit{Chlamydomonas} is trapped into perfect circular paths with radius $r$,  linearly proportional to the mean out-of-plane curvature $\kappa_1$, shown in Fig.~\ref{fig:chlam_fig}(d).
Interestingly, any incongruousness in the mean intrinsic curvature between the two flagella is sufficient to allow \textit{Chlamydomonas} to swim progressively via an array of helical paths in 3D, depending on differences in the orientation of their beating-planes, as observed experimentally \cite{cortese_control_2021,wilson_chiral_2022}.

By adding an asymmetry between the two flagella's out of plane curvatures, the circular paths seen in Fig.~\ref{fig:chlam_fig}(d) turn into helical paths shown in Fig.~\ref{fig:chlam_fig}(e). If instead the two out of plane curvatures are equal and opposite, the \textit{Chlamydomonas} swims along a straight trajectory and rolls about its own axis, shown in Fig.~\ref{fig:chlam_fig}(f). Again, breaking the symmetry between the two flagella produces more complex swimming. In the case of opposite sign and asymmetric out of plane curvatures, the trajectory of \textit{Chlamydomonas} becomes helical (distinct from the previous helical case) (Fig.~\ref{fig:chlam_fig}(g)), in further agreement with 3D observations \cite{cortese_control_2021,wilson_chiral_2022}.

\subsection{Sperm-egg elastohydrodynamic scattering}
The journey of sperm cells to the egg is one of the most important biological processes found in nature. The fertilization process involves the interaction of multiple bodies with varying size, active and passive components, and solid and elastic structures, providing endless opportunities biolgoical soft-matter and fluid-structure research \cite{gaffney_mammalian_2011}. Most of sperm swimming modelling focuses on hydrodynamic simulations using prescribed swimming beat patterns (kinematic constrained) to resolve swimming trajectories \cite{ishimoto_simulation_2016}. However, this bypasses the potential for elastohydrodynamic modulations in the system. In this example, we model for the first time the elastohydrodynamic scattering between a sperm and and egg freely floating in the fluid. The CG formalism allows investigation of elastohydrodynamic swimming modulation and simple construction of elastic and passive solid bodies, that are fixed or free to move in the fluid. The latter is particularly challenging given that free force/torque balance governs the movement of the body as a whole. Here, we briefly test the hypothesis that spherical egg alone is able to hydrodynamically attract the sperm to the egg via non-local hydrodynamic interactions, and flagella bending re-orientation, as have been observed for bacteria swimming near spherical obstacles \cite{takagi_hydrodynamic_2014,spagnolie_geometric_2015}.

 For simplicity, we model the sperm head with a spherical shape of radius $0.2L$ and the egg as a sphere of radius $1L$. We note that eggs vary in size depending on the species  \cite{ahlstrom_characters_1980}. The sperm flagella is made of $N = 16$ segments with $n = 1$ sphere per segment. Using an initial sperm-egg separation (measured from centre of sperm head to centre of egg) of $5L$, we vary $\theta$, the angle that the sperm cell makes with the $z$-axis at $t = 0$. Rather than prescribing a time-dependent intrinsic curvature, we set an active internal moment density along the flagella, $\mathbf{m} = 12k\cos{(ks-t)}\mathbf{d}_2$, that is integrated from $s_j$ to $L$ and added to bending moment at $s_j$. This causes the waveform shown in Fig.~\ref{fig:model_sperm}(a). Simulations are run with $\mathcal{S}= 6$.  Fig.~\ref{fig:model_sperm}(b) shows the initial condition for a single value of $\theta$, and the subsequent trajectories of sperm cells towards the egg for four values of $\theta$. The sperm cell that approaches closest to the egg shows the most noticeable curve in trajectory, as well as the largest change in speed as it swims past the egg (Fig.~\ref{fig:model_sperm}(c)). 
 
 Fig.~\ref{fig:model_sperm}(d) shows the sperm swimming speed as a function of sperm-egg separation; initially separated by 5 $(L)$, the sperms approach the egg and their swimming speed slows. This reduction in swimming speed is up to 28 \% for the sperm with the closest approach. Once the sperm reaches the closest separation from the egg, its swimming speed abruptly increases before gradually returning to its free-space swimming speed (i.e. the swimming speed in absence of solid bodies), as the sperm-egg separation increases. These simulations indicates that hydrodynamic effects alone are unable to assist sperm to find eggs with similar sizes as the sperm, even when swimming in close proximity. This may explain need for other form of guidance mechanisms through evolution, such as chemotaxis in freshwater fish species \cite{kholodnyy_how_2020}. Most interestingly, Fig.~\ref{fig:model_sperm}(d) seems to indicate the possibility of `accelerating' microorganisms by solid body scattering, though this depends on the proximity of the swimmer to the free body. The curves in Fig.~\ref{fig:model_sperm}(d) are skewed, and above the average speed for certain sperm-egg separations are permitted, so that the accumulated distance travelled is higher for those cases (see purple and yellow curves in Fig.~\ref{fig:model_sperm}(d)). Trajectories are qualitatively similar to trajectories presented in Ref.~\cite{ishimoto_simulation_2016}. In all, the exemplary results in Fig.~\ref{fig:model_sperm} motivate further studies on the design and optimization of micro-environments to promote, or suppress, cell transport and progression using microfluidic systems.

\begin{figure*}[t!]
    \includegraphics[width=\textwidth]{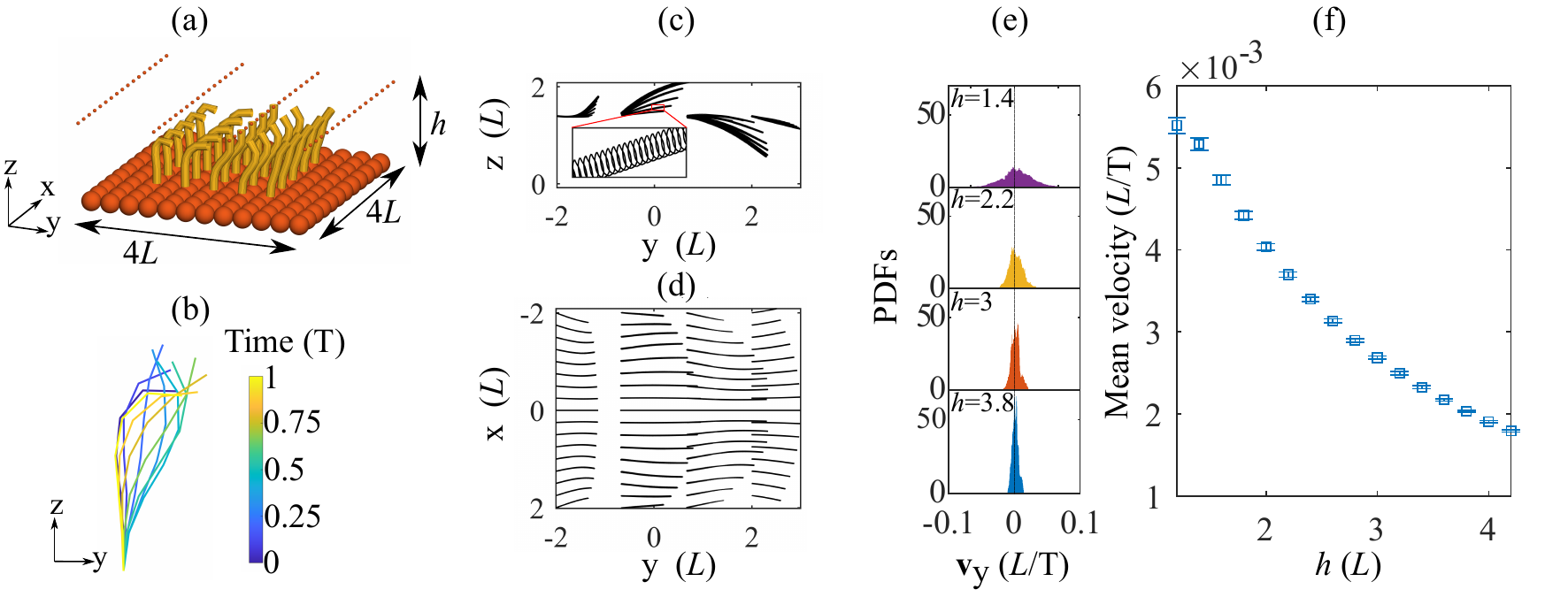}
    \caption{(a) Initial condition for $5\times5$ cilia array with wall made of spheres and tracer particles positioned above. Tracer particle radius $r = 0.01\ (L)$ (plotted larger for visual purposes). (b) Beating of a single cilium over one period with time progressing from black to white. Cilia beat in the $y-z$ plane and induce net fluid flow in the positive $y$ direction. The phase of cilia beating varies by $2\pi$ across the array, with cilia at the same $y$ having identical phase. Tracer particles pushed by fluid flow created by cilia array. (a) Birds-eye view of trajectory of tracer particles. Their direction of travel is from negative $y$ to positive $y$. (b) Side-view of trajectory of tracer particles. (c) PDF of tracer particle velocity in the $y$ direction, dependent on initial height. Higher particles are further from cilia array and as a result have lower average velocity in the $y$ direction (inset).}
    \label{fig:cilia}
\end{figure*}

\subsection{Elastohydrodynamic cilia array and particle transport}
Arrays of cilia are found in many places in our body and are responsible a wide array of biological processes, from driving fluid flow during embryonic growth to clearing mucus in the lungs \cite{cartwright_fluid-dynamical_2004}. Elastohydrodynamic simulations of cilia array have been developed for decades now to investigate several multi-scale aspects of this system, including studies on synchronization phenomena and metachronal waves \cite{gueron_cilia_1997,meng_conditions_2021, westwood_coordinated_2021,elgeti_emergence_2013,gilpin_multiscale_2020}.
Here, we consider this canonical system using the CG formalism for the first time. The method outlined here is capable of simulating arrays of cilia and tracking fluid flow, which we demonstrate with a simple example. We arrange $25$ cilia into a $5\times5$ grid arranged on top of a wall made of spheres, as shown in Fig.~\ref{fig:cilia}(a). Each cilia is made of $N = 7$ segments with $n = 1$ spheres per segment and governed by a prescribed curvature $\boldsymbol{\kappa} = (1 + 5\sin(1.5\pi s-t+\phi))\mathbf{d}_1$, Fig.~\ref{fig:cilia}(b). Each row (cilia with same location in $y$) in the cilia array has a different value for the phase $\phi$ such that a metachronal wave is travelling in the positive $y$ direction, Fig.~\ref{fig:cilia}(a). Specifically, $\phi = (i-1)/5 \times 2\pi$ where $i$ corresponds to the row number of the cilium, for $i = 1, ..., 5$. The beating of each individual cilium is shown in Fig.~\ref{fig:cilia}(b) over one period of oscillation. We note that one may use modified mobility tensors accounting for the hydrodynamics of a wall, but we choose to use spheres to model the wall for generality. In our example, the wall is fixed, but is easily made free, and can also be transformed into a non-planar ciliated surface (e.g. a curved wall) to investigate the effect of different wall geometries \cite{westwood_coordinated_2021}, something that is not immediately possible using modified hydrodynamic tensors.

The metachronal wave induces a net fluid flow above the cilia array which transports tracer particles positioned above the array. The resultant trajectory of tracer particles with an initial height of 1.4 $(L)$ can be seen in Fig.~\ref{fig:cilia}(c,d) from a birds-eye and side-view, respectively. The inset in Fig.~\ref{fig:cilia}(c) shows the oscillations of a single tracer particle over the beating period, and correlates with ciliary beating driving them. Over long time, the tracer particles are successfully transported to the positive side of the $y$ axis, in accordance with the biased waveform in Fig.~\ref{fig:cilia}(b), with particles near the centre of the array moving faster than those at the sides due to the finite size of the array. Other boundary effects of the patch of cilia can be seen in Fig.~\ref{fig:cilia}(c), in which particles tend to move upward (downward) for negative (positive) $y$. The PDF of particle velocity in the $y$ direction for different initial heights in Fig.~\ref{fig:cilia}(e) shows that the particle transport is not effective for this ciliary beating. The transport velocity is only weakly shifted to positive values due to small component of the beating in the direction of transport, with the average velocity of transport decaying with particle height, shown in Fig.~\ref{fig:cilia}(f). Distributions of accessible velocities by the tracer particles vary noticeably with particle height due to the fast spacial characteristic decay of the velocity field in low Reynolds number.

\section{Discussion and conclusions}
We offer a simple and intuitive method for simulating fully three-dimensional interacting elastic filaments coupled via non-local multi-body microhydrodynamics (Matlab code provided). By using asymptotic integration of momentum balance along each coarse-grained segment \cite{moreau_asymptotic_2018}, we avoid challenging numerical treatment of high-order fluid-structure interaction PDEs and the numerically stiff Lagrange multipliers required to enforce the inextensibility constraint. Our method allows straightforward construction of complicated multi-filament-body systems involving moving, fixed and coupled-filament conditions, allowing easier treatment of cumbersome force-free and torque-free constraints. Our CG formulation features the construction of complex solid body geometries using spheres as building blocks, and the exponential mapping of quaternions to overcome coordinate-singularities in 3D with a trivial rescaling.

The CG framework recasts the 3D non-local elastohydrodynamic PDEs into an ODE system of equations, bypassing altogether the need of numerically meshing partial derivatives. This allows the use of generic numerical solvers. As such, no experience is needed in numerical methods, and only basic knowledge of linear algebra and matrix manipulation are required to use this framework. The full elastohydrodynamic system is simply condensed into block-operators,
\begin{equation}
	\boldsymbol{\mathcal{M}}_F\boldsymbol{\mathcal{M}}_H^{-1}\boldsymbol{\mathcal{Q}}\boldsymbol{\mathcal{D}}\dot{\boldsymbol{\mathcal{X}}}_{gen} = \boldsymbol{\mathcal{K}},
	\label{sys_expmap_explained}
\end{equation}
 each encoding distinct model interactions of the shape field  $\dot{\boldsymbol{\mathcal{X}}}_{gen}$, namely: momentum balance $\boldsymbol{\mathcal{M}}_F$, hydrodynamic coupling $\boldsymbol{\mathcal{M}}_H^{-1}$, dimensional reduction $\boldsymbol{\mathcal{Q}}$, geometry of deformation and basis tracking $\boldsymbol{\mathcal{D}}$, and constitutive relations, internal activity and other boundary constraints $\boldsymbol{\mathcal{K}}$. Eq.~(\ref{sys_expmap_explained}) is succinct and facilitates model explainability, generalisations and/or simplifications, as well as analytical and numerical analysis, better treatment of boundary conditions, and easier implementation in different platforms. The formalism allows model customization without the need of algorithmic or numerical re-design. For example, the reduction in dimensionality operator may be removed if needed, the non-local hydrodynamic block may be replaced by other local or non-local drag theories, or by numerical solutions from a Navier-Stokes solver (e.g. immersed boundary methods). The exponential mapping of quaternions may be replaced by other coordinate parametrizations of choice and different material properties of the filament may be equally invoked. Other types of forces arising from, for example, gravity, magnetism, electrostatic and stochastic interactions, are straightforward to implement, offering further flexibility whilst employing this methodology.

The method has been validated and contrasted against previous experimental and computational studies showing excellent agreement with the literature (Fig.~\ref{validations}). We show that the exponential mapping parametrization provides a simpler implementation, and generally faster running times, than a direct quaternion implementation, but both parametrizations are robust while tracking multiple interacting filaments. Finally, we showcase the method in three complex elastohydrodynamic examples: (1) bi-flagellated {\it Chlamydomonas} swimming in 3D (Fig.~\ref{fig:chlam_fig}), (2) sperm-egg elastohydrodynamic coupling (Fig.~\ref{fig:model_sperm}), and (3) canonical particle transport by cilia array (Fig.~\ref{fig:cilia}). The elastohydrodynamic models and results presented for examples (1) and (2) are novel. Indeed, the computational realization of these systems using the classical PDE formulation is still challenging today, specifically regarding the implementation of the boundary conditions and other filament constraints. 

Our simulations revealed that the non-local hydrodynamic coupling increases the elastohydrodynamic propulsion of the \textit{Chlamydomonas} swimming. Most importantly, the 3D elastohydrodynamic model is able to predict a bewildering array of complex 3D swimming trajectories that may arise via simple symmetry-breaking features of the flagella beat in 3D, from circular to helical trajectories (Fig.~\ref{fig:chlam_fig}), in agreement with~\cite{cortese_control_2021,wilson_chiral_2022}. This offers numerous modelling opportunities motivated by recent 3D observations of this important model microorganism \cite{cortese_control_2021,wilson_chiral_2022}. We also report novel sperm-scattering results in which sperm swimming can either be enhanced or reduced by the presence of a free-body (Fig.~\ref{fig:model_sperm}). Finally, we observe canonical small-scale oscillations of tracer particles in synchrony with ciliary beating, and show that the non-local particle transport decays with height (Fig.~\ref{fig:cilia}), thus mimicking generic properties of ciliary systems, for the first time using the CG framework.

Numerous open questions exist in elasto-microhydrodynamics systems in which this method could be readily employed: self-organisation of flagellar beat~\cite{oriola_nonlinear_nodate,chakrabarti_spontaneous_2019}, elastohydrodynamic modulation of active bundles \cite{pochitaloff_flagella-like_2022}, non-trivial body geometries \cite{wang_self-adaptive_2021}, physicochemical active guidance \cite{kholodnyy_how_2020},  ciliated microorganisms \cite{drescher_dancing_2009}, cilia carpets \cite{wang_cilia_2022}, artificial swimmers \cite{magdanz_ironsperm_2020,magdanz_development_2013}, soft-robotics \cite{milana_metachronal_2020,li_supramolecularcovalent_2020}, locomotive robotics \cite{renda_geometric_2018} and the role of torsional instabilities in filament systems \cite{chakrabarti_flexible_2020}, to name a few. Extending the framework to include inertial forces or the full Cosserat rod Theory would widen the scope for the method's use. In all, we hope that the ease-of-implementation advantages of the method will facilitate the study of non-trivial fluid-structure interaction systems, and assist researchers to fast-track implementation and accelerate the modelling cycle, serving communities within, and away from, mathematical and physical sciences that are equally interested in simulating these systems.

\section*{}
\noindent{{\textbf{Data accessibility.}}} The MATLAB codes used to run the simulations presented in this paper are available on github via the link:\\
\textbf{Funding.} Funded by DTP-EPSRC. \\
{{\textbf{Competing interests.}} We declare we have no competing interests.}\\
{{\textbf{Acknowledgements.}} This work was carried out using the computational facilities of the Advanced Computing Research Centre, University of Bristol - http://www.bristol.ac.uk/acrc/.}

\appendix
{
\section{Matrix cross product}
\label{appen_cross}
The cross product $\mathbf{a}\times\mathbf{b}$ in matrix form is given by
\begin{equation}
		\mathbf{a} \times \mathbf{b} = [\mathbf{a}]_\times \mathbf{b} = \begin{bmatrix}
			0 & -a_3 & a_2 \\
			a_3 & 0 & -a_1 \\
			-a_2 & a_1 & 0
		\end{bmatrix}
		\begin{bmatrix}
			b_1 \\ b_2 \\ b_3
		\end{bmatrix}.
	\end{equation}

	\section{Force and torque balance of two free filaments} \label{appendix:two_free_filaments}
For a single filament with no body, the block of zeros in the large matrix in Eq.~\ref{Mf_fil_body} reduces to zero size, and the number of spheres in the system is given by $M = Nn$. If $\mathbf{M}_F$ encodes the force and torque balance of a single filament, e.g.
\begin{equation}
	\mathbf{M}_F \begin{bmatrix}
		\boldsymbol{\mathcal{F}} \\ \boldsymbol{\mathcal{T}}
	\end{bmatrix} = \boldsymbol{\mathcal{K}},
\end{equation}
then the force and torque balance of two free filaments is given by
	\begin{equation}
	\begin{bmatrix}
		(\mathbf{M}_F)_1 & \mathbf{0} \\
		\mathbf{0} & (\mathbf{M}_F)_2
	\end{bmatrix}  \begin{bmatrix}
		\boldsymbol{\mathcal{F}}_1 \\ \boldsymbol{\mathcal{T}}_1 \\ 	\boldsymbol{\mathcal{F}}_2 \\ \boldsymbol{\mathcal{T}}_2
	\end{bmatrix} = \begin{bmatrix}
		\boldsymbol{\mathcal{K}}_1 \\ \boldsymbol{\mathcal{K}}_2
	\end{bmatrix},
\label{two_fil_example_Mf}
\end{equation}
where $(\mathbf{M}_F)_i$, $\boldsymbol{\mathcal{F}}_i$, $\boldsymbol{\mathcal{T}}_i$ and $\boldsymbol{\mathcal{K}}_i$ have identical structure to the single filament case and $i = 1, 2$ denotes the filament number. When writing in general form 
\begin{equation}
\boldsymbol{\mathcal{M}}_F \begin{bmatrix}
	\boldsymbol{\mathcal{F}} \\ \boldsymbol{\mathcal{T}}
\end{bmatrix} = \boldsymbol{\mathcal{K}},
\end{equation}
rows and columns in Eq.~\ref{two_fil_example_Mf} are rearranged accordingly. Adding additional structures results adding extra matrices along the diagonal in Eq.~\ref{two_fil_example_Mf}. For example, adding a single free sphere to the system would add 6 additional rows to $\boldsymbol{\mathcal{M}}_F$, corresponding to one identity matrix multiplying the force experienced by the sphere, and one identity matrix multiplying the torque experienced by the sphere. Six extra zeros would be added to the right-hand-side corresponding to the force- and torque-free conditions.
	\section{Hydrodynamic mobility matrix}
	\label{Hydro-appen}
	Using the Rotne-Prager-Yamakawa approximation gives the hydrodynamic mobility matrix
\begin{equation}
	\boldsymbol{\mathcal{M}}_H = \begin{bmatrix}\boldsymbol{\mu}^{tt}  & \boldsymbol{\mu}^{tr} \\ \boldsymbol{\mu}^{rt}& \boldsymbol{\mu}^{rr}\end{bmatrix},
\end{equation}
where $t$ and $r$ refer to translational and rotational components, respectively. The submatrices have general form 
\begin{align}
	\boldsymbol\mu = \begin{bmatrix}
		\mu_{11} & \mu_{12} & \hdots & \mu_{1N_{sph}} \\
		\mu_{21} & \mu_{22} & \hdots & \mu_{2N_{sph}} \\
		\vdots & \vdots & \ddots& \vdots \\		
		\mu_{N_{sph}1} & \mu_{N_{sph}2} & \hdots & \mu_{N_{sph}N_{sph}} 
	\end{bmatrix}.
\end{align}
Components of $\boldsymbol{\mu}^{tt}, \boldsymbol{\mu}^{tr},\boldsymbol{\mu}^{rt}$ and $\boldsymbol{\mu}^{rr}$ are given by \cite{zuk_rotneprageryamakawa_2014}
\begin{align}
	\mu_{ij}^{t t}= \begin{cases}\frac{1}{6 \pi \eta a_{i}} \mathbf{I} & i=j, \\ \frac{1}{8 \pi \eta r_{ij}}\left(\left(1+\frac{a_{i}^2+a_{j}^2}{3 r_{ij}^{2}}\right) \mathbf{I}+\right.\\ \left.\left(1-\frac{a_{i}^2+a_{j}^2}{r_{ij}^{2}}\right) \hat{\boldsymbol{r}}_{ij} \hat{\mathbf{r}}_{ij}\right) & i \neq j,\end{cases}
\end{align}
\begin{equation}
	\mu_{ij}^{r r}= \begin{cases}\frac{1}{8 \pi \eta a_{i}^{3}} \mathbf{I} & i=j, \\ \frac{1}{16 \pi \eta r_{ij}^{3}}\left(3 \hat{\boldsymbol{r}}_{ij} \hat{\boldsymbol{r}}_{ij}-\mathbf{I}\right) & i \neq j,\end{cases}
\end{equation}
\begin{equation}
	\mu_{ij}^{t r}=\mu_{ij}^{r t}= \begin{cases}0 & i=j, \\ \frac{1}{8 \pi \eta r_{ij}^{2}} \boldsymbol{\varepsilon} \cdot \hat{\boldsymbol{r}}_{ij} & i \neq j,\end{cases}
\end{equation}
where $a_i$ is the radius of sphere $i$, $\eta$ is the fluid viscosity, $\mathbf{r}_{i,j}$ is the separation of sphere $i$ and $j$ and $\boldsymbol{\varepsilon}$ is the Levi-Civita symbol. When $i=j$, $\mu^{tt}_{ij}$ and $\mu^{rr}_{ij}$ reduce to standard force and torque relation for a sphere in the Stokes regime, and there is no translation-rotation component. For $i \neq j$, the tensors dictate how sphere $j$ affects the movement of sphere $i$. For example, $\mu^{tt}_{ij}$ couples the force on sphere $j$ with the velocity of sphere $i$; $\mu^{tr}_{ij}$ couples the torque on sphere $j$ with the velocity of sphere $i$ and $\mu^{rr}_{ij}$ couples the torque on sphere $j$ with the angular velocity of sphere $i$. The tensors used here allow for spheres of different size, allowing different structures to be constructed out of differently sized spheres, depending upon the precision required.  

\section{Reducing dimensionality of system}
\label{reduction_dimension}
	Consider a single filament attached to a solid body. Spheres are rigidly attached within each segment, and the velocity at $\mathbf{x}_i$ is given by the velocity at $\mathbf{x}_{i-1}$ plus a contribution from the angular velocity of segment $i-1$. Additionally, the velocity of spheres in the body is given by the velocity $\dot{\mathbf{x}}_b$ plus a contribution from the angular velocity of the body $\boldsymbol{\omega}_b$ As such, the velocity of a sphere in the system is given by
\begin{widetext}
\begin{equation}
	\mathbf{v}_i = \begin{cases}
	\dot{\mathbf{x}}_b - \boldsymbol{\Delta}_i \times \boldsymbol{\omega}_b & i\leq N_{body},\\
\dot{\mathbf{x}}_b-(\mathbf{x}_1 - \mathbf{x}_b)\times \boldsymbol{\omega}_b - \Delta s \sum_{j=1}^{\ceil{k/n}-1} \mathbf{d}_3^j\times \boldsymbol{\omega}_j^{seg} - \mathbf{d}^{\ceil{k/n}}_3 \times \boldsymbol{\omega}_{\ceil{k/n}}^{seg}  & i > N_{body},
	\end{cases} \\
\label{VQ}
\end{equation}
\end{widetext}
where $\Delta s$ is the length of each segment, $k = i-N_{body}$, $\ceil{k/n}$ rounds $k/n$ up to the nearest integer and $\boldsymbol{\omega}_i^{seg}$ is the angular velocity of the $i\textsuperscript{th}$ segment in the filament. The quantity $\ceil{k/n}$ gives the segment number that sphere $i$ is attached to. The angular velocity of a sphere in the system is given by
\begin{equation}
		\boldsymbol{\omega}_i  =\begin{cases}
			\boldsymbol{\omega}_b & i\leq N_{body},\\
			\boldsymbol{\omega}_{\ceil{k/n}}^{seg}  & i > N_{body}.
		\end{cases}
	\label{omegaQ}
\end{equation}
Equations (\ref{VQ}) and (\ref{omegaQ}) allow all $\mathbf{v}_i$ and $\boldsymbol{\omega}_i$ in the system to be written in terms of velocity of the centre of mass of the body $\dot{\mathbf{x}}_b$, its angular  velocity $\boldsymbol{\omega}_b$ and the angular velocity of each filament segment  $\boldsymbol{\omega}_i^{seg}$. By additionally rigidly attaching the first segment of the filament to the body, the dimensionality of the system is reduced from $6(N_{body}+Nn)$ to $3+3N$ velocities. Writing equations (\ref{VQ}) and (\ref{omegaQ}) in matrix form gives Eq.~(\ref{Q_1}) in the main text.	
	}

\section{Implementing quaternion parametrisation} \label{appendix:quaternions}
Consider again the single filament attached to a body. Assign a quaternion $\mathbf{q}_b$ to the body frame vectors and $\mathbf{q}_i$ to filament segment frame vectors. The state vector for the system is
\begin{equation}
	\mathbf{X}_{quat} = \begin{bmatrix}
		\mathbf{x}_b^\top & \mathbf{q}_b^\top& \mathbf{q}_1^\top & \hdots & \mathbf{q}_N^\top
	\end{bmatrix}^\top.
\end{equation}
The rate of change of the state vector is given by
\begin{equation}
	\dot{\mathbf{X}}_{quat} = \begin{bNiceMatrix}
		\mathbf{I}  & \mathbf{0}& \Cdots  & &  &\mathbf{0} \\
		\mathbf{0} & \mathbf{P}(\mathbf{q}_b) & \mathbf{0} & \Cdots &   &\mathbf{0}  \\
		\mathbf{0} & \mathbf{0} & \mathbf{P}(\mathbf{q}_1) & \mathbf{0} & \Cdots   &\mathbf{0}  \\
		\Vdots &  & & &     & \Vdots  \\
		\mathbf{0} & \Cdots&  &  &  \mathbf{0} &\mathbf{P}(\mathbf{q}_N)  \\
		\CodeAfter \line{3-3}{5-6} 
	\end{bNiceMatrix}
	\begin{bNiceMatrix}
		\dot{\mathbf{x}}_b \\ \boldsymbol{\omega}_b \\ \boldsymbol{\omega}_1^{seg} \\ \vdots \\ \boldsymbol{\omega}_N^{seg}
	\end{bNiceMatrix}
 \label{C_mat_ex}
\end{equation}
where $\mathbf{P}(\mathbf{q})$ is given by (see Eq. \ref{qi_omega})
\begin{equation}
	\mathbf{P}(\mathbf{q}) = \frac{1}{2}\begin{bmatrix}
		-q_1 & -q_2 & -q_3 \\
		q_0 & q_3 & -q_2 \\
		-q_3 & q_0 & q_1 \\
		q_2 & -q_1 & q_0
	\end{bmatrix}.
\end{equation}
By rigidly attaching the first segment of the filament to the body, $\boldsymbol{\omega}_1^{seg} = \boldsymbol{\omega}_b$, the dimension of the vector on the RHS of Eq.~\ref{C_mat_ex} can be reduced by 3, in addition to removing one column from the matrix pre-multiplying the vector. Generalising to arbitrary systems gives the dynamics of the state vector $\boldsymbol{\mathcal{X}}_{quat}$ in terms of the reduced velocity system $\boldsymbol{\mathcal{W}}$ via the matrix $\boldsymbol{\mathcal{C}}$,
\begin{equation}
	\dot{\boldsymbol{\mathcal{X}}}_{quat} = \boldsymbol{\mathcal{C}}\boldsymbol{\mathcal{W}}.
	\label{quat_timestep}
\end{equation}

	\bibliography{refs_final}
	\bibliographystyle{vancouver}	
\end{document}